\documentclass{article}
\usepackage{times}
\usepackage{natbib}
\usepackage{graphicx}
\usepackage[a4paper]{geometry}
 
\begin{document}

\def\mnras{MNRAS}
\def\aj{AJ}
\def\aap{A\&A}
\def\apj{ApJ}
\def\nat{Nature}
\def\apjl{ApJ}
\def\pasp{PASP}
\def\apjs{ApJ}

\title{The Kinematic Age of the Coolest T Dwarfs} 
\author{L. Smith$^1$, P. Lucas$^1$, B. Burningham$^1$, H. Jones$^1$, R. Smart$^2$, D. Pinfield$^1$, \\F. Marocco$^1$, and J. Clarke$^1$\\
                {\small{\it $^1$ Centre for Astrophysics Research, Science and Technology Research Institute, University of Hertfordshire, HatÞeld AL10 9AB}}\\
                {\small{\it$^2$ Istituto Nazionale di AstroÞsica, Osservatorio Astronomico di Torino, Strada Osservatrio 20, 10025 Pino Torinese, Italy}}\\
                {\small\texttt{l.smith10@herts.ac.uk}}\\
                }
\date{Article version of a poster paper at Cool Stars 17, Barcelona, June 2012} 
\maketitle
 
\begin{abstract}
Surprisingly, current atmospheric models suggest that the coolest T dwarfs (T8.5 to T10) are young and very low mass ($0.06-2Gyr$, $5-20M_{jup}$, \citealt{leggett09}, \citeyear{leggett10}, \citeyear{leggett12}). Studies of population kinematics offer an independent constraint on the age of the population. We present kinematic data of a sample of 75 mid to late T dwarfs drawn from a variety of sources. We define our samples, T5.5 to T8 and T8.5 to T10, as mid and late T respectively. UKIDSS LAS kinematics were derived from our automated LAS proper motion pipeline and distance estimates derived from spectral types and photometry for the minority of sources that lack parallaxes. Our results show that the mid and late T populations do not have distinctly separate tangential velocity distributions to 95\% probability. They also give an approximate mean kinematic age equal to that of a population with $B-V$ colour $0.51-0.54$, and a spectral type late F, which corresponds to an age of about $2Gyr$. However the median and modal ages are greater. This indicates that while model atmospheres correctly predict some trends in colour with gravity and age, reliable ages cannot yet be inferred from them. More benchmark objects are needed to anchor the models.
\end{abstract}
 
\section{Background and Aims}
Current atmospheric models suggest that the coolest T dwarfs from the UKIDSS survey are young and very low mass. For example the T8.5 dwarf ULAS 1238 has a model mass and age of $6-10 M_{jup}$ and $0.2-1Gyr$, T9 dwarfs ULAS 0034 and ULAS 1335 both have model masses and ages of $5-20 M_{jup}$ and $0.1-2 Gyr$ (\citealt{leggett09}, \citeyear{leggett10}), and T10 dwarf ULAS 0722 has a model mass and age of $3-11 M_{jup}$ and $0.06-1Gyr$ \citep{leggett12}. This is surprising, theory predicts that these objects cool at a steady rate through to later types. Young, planetary mass late T dwarfs should be very rare unless the Initial Mass Function rises at planetary masses. Studies of population kinematics offer an independent test of the age of a population. We present results of a method to determine the kinematic age of mid-late T dwarfs, in an effort to test the models. 

\begin{figure}[ht]
\centering
\includegraphics[width=0.8\textwidth,natwidth=1181,natheight=479]{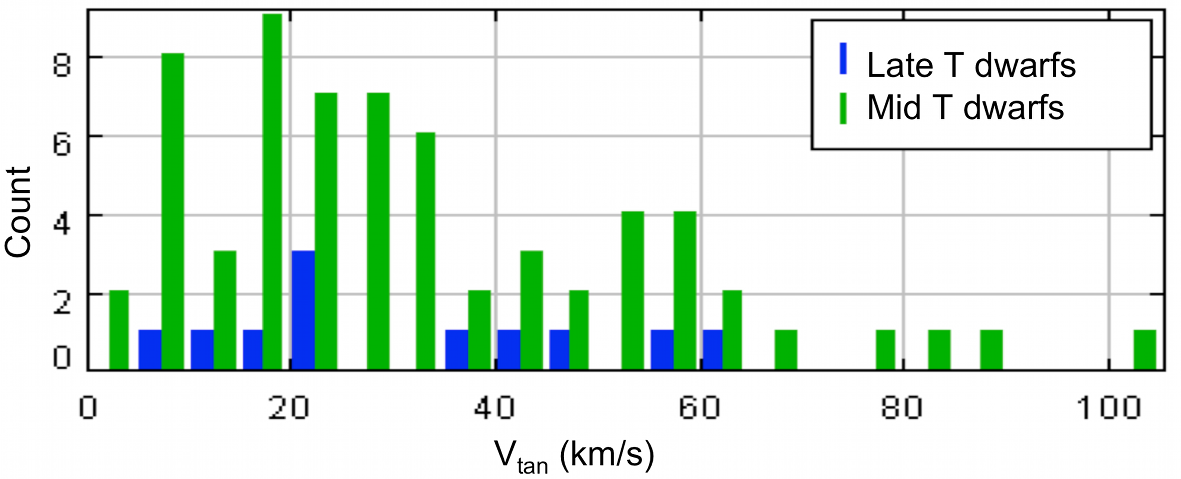}
\caption{The distribution of LSR-corrected tangential velocities for the mid and late T dwarf samples. T5.5 to T8 inclusive is considered a mid T dwarf, T8.5 and onwards is considered a late T dwarf. The histogram suggests the V$_{tan}$ distributions do not differ significantly.}
\label{vtanhist}
\end{figure}

\section{Our Kinematic Sample}
Our sample is drawn from the UKIDSS late T dwarf sample and \citet{dupuy12}. Parallaxes are available from the PARSEC parallax program (\citealp{marocco10}; \citealp{andrei11}), \citet{dupuy12} and \citet{kirkpatrick12} for 48/75 T dwarfs. Distance estimates for the remainder are derived from photometry and spectral types using the polynomial fits of J band absolute magnitude to spectral type described in \citet{marocco10}. Proper motions are taken where given by the sources of the parallax measurements and from our bespoke UKIDSS LAS proper motion pipeline otherwise. Proper motions are also LSR corrected by us.

\begin{figure}[ht]
\centering
\includegraphics[width=\textwidth,natwidth=1181,natheight=425]{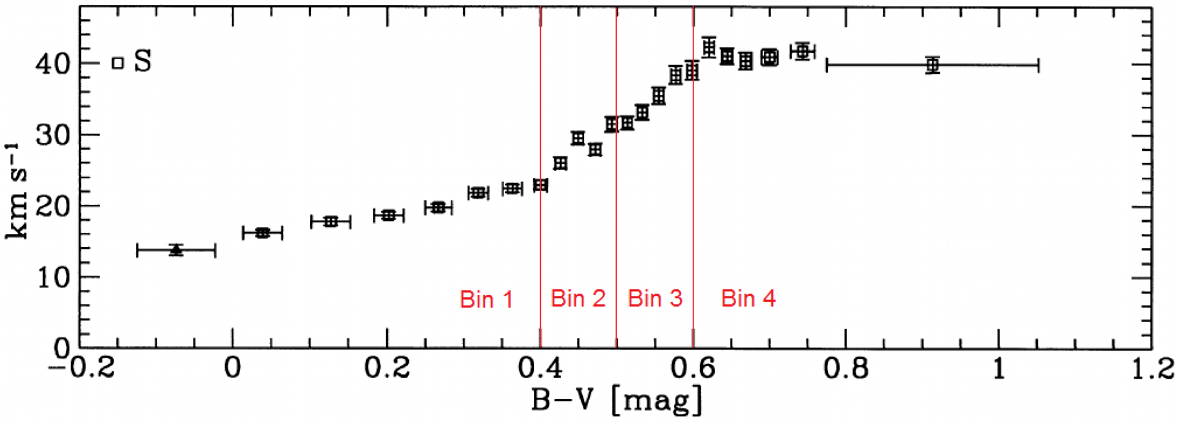}
\caption{Total velocity dispersion of main sequence Hipparcos stars vs. $B-V$ colour. Our binned sections are also shown for reference. Figure adapted from Dehnen \& Binney (1998). Our result suggests that the samples of mid and late T dwarfs are consistent with a kinematic dispersion similar to that of a mid to late F type star, $B-V$ of about 0.5.}
\label{bminvplot}
\end{figure}

\section{Methods}
\begin{figure}[hb]
\centering
\includegraphics[width=1.1\textwidth,natwidth=1181,natheight=408]{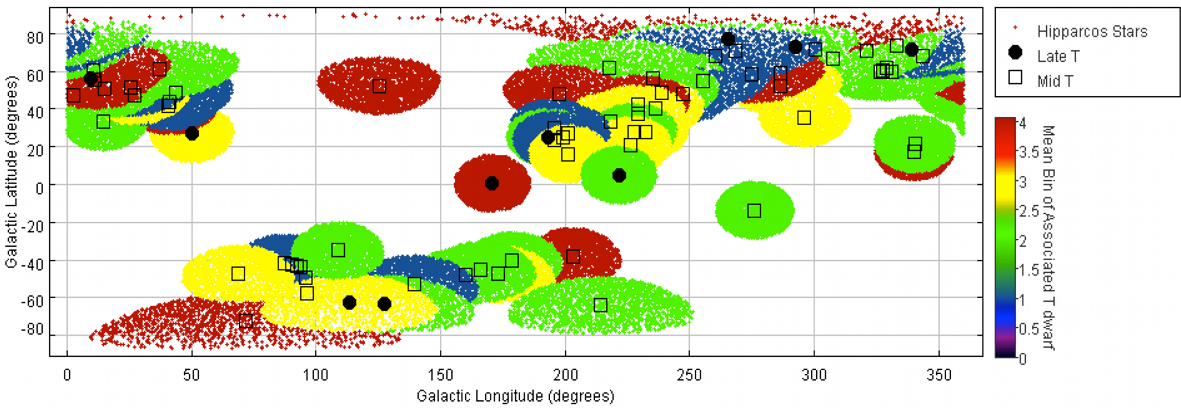}
\caption{The distribution of the sample of T dwarfs, each is surrounded by a coloured patch showing the positions of the Hipparcos stars used for tangential velocity comparison. The dwarfs are reasonably well distributed in longitude but the galactic plane is relatively poorly sampled. The colours show positions of the Hipparcos  stars used for comparison to different dwarfs.}
\label{areaplot}
\end{figure}

We performed two-sample K-S and StudentÕs t tests on the total tangential velocities of mid and late T dwarf populations (totalling 64 and 11 dwarfs respectively). 
We then compared the RA and Dec components of V$_{tan}$ for each T dwarf from our sample to the median V$_{tan}$ components of Hipparcos stars within 15 degrees on the sky (see Figure \ref{areaplot}). The purpose of the radial selection criteria was to remove possible differential contributions to V$_{tan}$ based on galactic coordinates. We used only Hipparcos stars with reliable parallax and genuine B-V measurements ($\frac{\sigma_{\pi}}{\pi} \leq 0.1$, $\sigma_{B-V} > 0$). The Hipparcos stars were separated into the 4 fairly broad B-V colour bins shown in Table \ref{bintable}.

\begin{table}
\centering
\begin{tabular}{|c|c|l|}
\hline
    \multicolumn{1}{|c|}{Bin}&
    \multicolumn{1}{c|}{Boundaries}&
    \multicolumn{1}{c|}{Typical Contents}\\
\hline
    1&$-\infty$ $<~B-V~<$ $0.4$&O, B, A, and early F\\
    2&$0.4$ $<~B-V~<$ $0.5$&Mid F\\
    3&$0.5$ $<~B-V~<$ $0.6$&Late F and early G\\
    4&$0.6$ $<~B-V~<$ $\infty$&Mid G onwards\\
\hline
\end{tabular}
\label{bintable}
\end{table}

Bin 1 encompasses 0, B, A and early F type stars which have relatively small total velocity dispersions due to their young ages. Bins 2 and 3 encompasses mid F to early G type stars with total velocity dispersion increasing fairly linearly with $B-V$ magnitude through the bin. Bin 4 contains stars with spectral types from mid G onwards, their main sequence lifetimes are comparable to the age of the Galaxy and as such their total velocity dispersions are reasonably constant. See Figure \ref{bminvplot}.
Our code then assigns a bin number to each component of tangential velocity for each T dwarf, the bin with the smallest difference in T dwarf velocity and median bin velocity being selected. The result is the mean bin number assigned across RA and Dec components of tangential velocity from each T dwarf population. 

\section{Results}
The K-S and StudentÕs t tests indicate that there is roughly a 95\% probability that the samples are drawn from the same population (K-S: 94\%, t: 97\%).

Results of the mid-late T dwarf to Hipparcos tangential velocity comparison give the two populations average colour bins of 2.6 and 2.9 (mid and late Ts respectively) showing that both populations have a kinematic dispersion similar to that of populations with $B-V$ colours of 0.51 and 0.54 (mid and late Ts respectively). It is worth noting that while the mean values of the selected bins are 2.6 and 2.9, the modal selected bin is 4 for both samples. Therefore the $\approx2Gyr$ characteristic age estimate based on the mean bin values may be an underestimate.

\section{Conclusion}
Our results show that there is no difference in the distributions of V$_{tan}$ of the late T and mid T dwarf populations, in contrast to what the models predict. A kinematic dispersion equal to that of a $B-V = 0.51-0.54$ colour star suggests a total velocity dispersion of order $38km~s^{-1}$ (with no $|W|$ velocity weighting applied). This corresponds to a stellar age of $\approx2Gyr$ (\citealp{binney00}; \citealp{fuchs01}). Median and modal ages are greater, of order $5Gyr$. $2Gyr$ is at the high end of ages predicted by models for ULAS 0034 and ULAS 1335 and older than predicted for ULAS 1238 and ULAS 0722. Thus the average ages predicted by the evolutionary models and kinematic measurements differ. This indicates that while model atmospheres predict some trends in colour with gravity and age well, reliable ages cannot yet be inferred from them. Increasing the number of known benchmark objects is necessary to anchor the current models.

\end{document}